\documentclass[12pt,cites,graphicx]{article}

\usepackage{geometry}                
\usepackage{graphicx}
\usepackage{amsmath}
\usepackage[all]{xy}
\usepackage{float}
\usepackage{stmaryrd}
\usepackage{amssymb}
\usepackage{amsfonts}
\usepackage{textcomp} 
\usepackage{verbatim}
\usepackage{graphicx}
\usepackage{epsfig}
\usepackage{epstopdf}

\newcommand{\ud}{\,\mathrm{d}}

\title{Two-dimensional electrochemical model for mixed conductors: a study of ceria}

\author{Francesco Ciucci,$^a$ \\ William C. Chueh$^b$ \\ Sossina M. Haile$^{b}$\\David G. Goodwin$^{a}$
 \\[3mm]
$^a$ California Institute of Technology \\
Department of Mechanical Engineering \\
Pasadena, CA USA. E-mail: frank@caltech.edu\\[1mm]
$^b$ California Institute of Technology \\ 
Department of Materials Science, \\
Pasadena, CA USA. E-mail: smhaile@caltech.edu}

\begin{document}

\maketitle

\renewcommand{\thefootnote}{\fnsymbol{footnote}}

\noindent
A two-dimensional small bias model has been developed for a patterned metal current collector $|$ mixed oxygen ion and electronic conductor (MIEC) $|$ patterned metal current collector electrochemical cell in a symmetric gas environment. Specifically, we compute the electrochemical potential distributions of oxygen vacancies and electrons in the bulk and near the surface for $\text{Pt} | \text{Sm}_{0.15}\text{Ce}_{0.85}\text{O}_{1.925} | \text{Pt}$ symmetric cell in a $\text{H}_2-\text{H}_2\text{O}-\text{Ar}$ (reducing) atmosphere from $500$ to $650^o C$. Using a two-dimensional finite-element model, we show that two types of electronic current exist within the cell: an in-plane drift-diffusion current that flows between the gas $|$ ceria chemical reaction site and the metal current collector, and a cross-plane current that flows between the two metal electrodes on the opposite side of the cell. By fitting the surface reaction constant $\tilde k_f^0$ to experimental electrode resistance values  while fixing material properties such as bulk ionic and electronic equilibrium defect concentrations and mobilities, we are able to separate the electrode polarization into the surface reaction component and the in-plane electron drift-diffusion component. We show that for mixed conductors with a low electronic conductivity (a function of oxygen partial pressure) or a high surface reaction rate constant, the in-plane electron drift-diffusion resistance can become rate-limiting in the electrode reaction. 

\section{Introduction}
Mixed ionic and electronic conductors have received significant attentions for their potential as fuel cell components, permeation membranes, oxygen storage capacitors, electrochemical sensors, etc. Electrical d.c. and a.c. conductivity measurements of the bulk focused mostly on separating the ionic and electronic contributions to the electrical conductivity as well as determining the dielectric and chemical capacitances. Physically derived one-dimensional models have aided in the interpretation of electrical conductivity data in the bulk and materials defect chemistry. On the other hand, investigations of interfaces in mixed conductors, specifically the gas $|$ mixed conductor and the metal $|$ mixed conductor interface, require a two-dimensional model to adequately describe the system due to nonlinearity of the electrochemical potential lines near the interface. However, the majority of the work in the community employs the one-dimensional treatment \cite{ISI:000168035300017},  \cite{Jamnik19994139} and only a handful of works attempted to scale up to two dimensional models, \cite{mebane:A421} \cite{Fleig:July2004:1385-3449:637} and \cite{Adler200035}. In this work, we developed a two-dimensional small bias model for the patterned metal current collector $|$ mixed conductor $|$ patterned metal current collector cell. In particular, we focus on the numerical analysis of the cross-plane electronic current that flows through the mixed conductor between patterned metal stripes on both side of the cell, and the in-plane electronic current that flows between the gas $|$ mixed conductor interface and the metal. In order for an electrochemical reaction to occur on the gas $|$ mixed conductor interface, electrons need to diffuse from the reaction site to the external circuit and vice-versa. Such a step, termed in-plane electron drift-diffusion, could play a significant role in the interfacial behavior of mixed conductors, particularly for those exhibiting a low to moderate bulk electronic conductivity. 

For this study, we have selected $\text{Pt} | \text{Sm}_{0.15}\text{Ce}_{0.85}\text{O}_{1.925} (SDC) | \text{Pt}$ as mixed oxygen ion and electron conductor model system. High oxygen ion conductivity of acceptor-doped ceria at intermediate temperatures ($500 -700^o C$) has attracted a great deal of interest in the solid oxide fuel cell (SOFC) community. In addition, under mildly reducing condition, doped ceria exhibits a moderate electronic conductivity ($\approx 0.1 S/cm$ at $650^oC$, $p_{O_2} =10^{-25} atm$ \cite{ISI:000232773100001}), making it attractive for fuel cell anode applications. Recent studies have also shown that, when operating SOFCs on hydrocarbon gases, ceria-based anode is significantly less susceptible to carbon coking \cite{Gorte00}. Insight into the in-plane electron diffusion path in ceria could lead to improved designs of anode geometries and reduced interfacial resistance.

\label{intro}

\section{Background}
The physical model, depicted in Fig. 1a, consists of a mixed oxygen ion and electron conductor (ceria) with patterned metal current collectors (Pt) on both sides placed in a uniform gas environment ($\text{H}_2-\text{H}_2\text{O}-\text{Ar}$ mixture). The patterned metal current collectors permit the system to be reduced to a repeating cell (Fig. 1b) using mirror symmetry lines ($\Gamma_1$, $\Gamma_2$, $\Gamma_3$). The thickness of the cell is given by $2l_2 = 1 mm$, while the surface dimensions are $2W_1\approx 3 \mu m $, the width of the metal $|$ ceria interface ($\Gamma_4$), and $2W_2 \approx 5 \mu m$, the width of the gas $|$ ceria interface ($\Gamma_5$) (Fig. 2). Two charge carriers species are considered: oxygen vacancies, denoted by the subscript Ò$ion$Ó, and electrons, denoted by Ò$eon$.Ó We solve the electrochemical potential and current of both charge carriers using a linear and time-independent model. 

We assume that the gas $|$ ceria interface is the prevailing surface active site facilitating the reaction between electrons and oxygen vacancies in the oxide and the gas phase species. In other words, the gas $|$ metal $|$ ceria triple-phase boundary interface has a negligible contribution to the surface reaction. As shown later in this work, mixed conductivity allows electrochemical reactions to take place away from the triple-phase boundary. As a result, the metal $|$ gas $|$ ceria interface, a 1D line, has substantially less area for reaction compared to the gas $|$ ceria interface, a 2D area. We further treat the surface chemistry as one global reaction, and do not consider diffusion of adsorbed species on the surface. Combined with the final assumption that the metal $|$ ceria interface is reversible to electrons, we are only considering two steps in the electrode reaction pathway: the surface reaction, and the electron drift-diffusion from the active site to the metal current collector. 

\label{intro}

\section{Model}
\subsection{Governing Equations}
Mixed-valent oxides such as SDC exhibit mixed oxygen ionic and n-type electronic conductivity behavior under reducing conditions. Oxygen ion motion is attributed to vacancy hopping while electron motion is attributed to thermally activated small polaron hopping (\cite{ISI:A1975V426800018} and \cite{Tuller-81}).  This complex behavior of ceria at the atomic scale can be averaged out at the mesoscale ($5 nm$ and up) leading to constitutive equations drawn from non-equilibrium thermodynamics \cite{PhysRevB.52.9406} and \cite{PhysRev.127.1004}). For electrons and oxygen vacancies we can write the following continuity equation:
\begin{equation}
{\partial{c_m}\over{\partial t}}+\nabla\cdot \mathbf j_m=\dot{\Omega}_m
\label{eqn:charge_carrier_transport}
\end{equation}
where $c_m$ is the concentration of charge carriers $m$ expressed in terms of number of particles per unit volume, $ \mathbf j_m$ is the particles mass flux, and $\dot{\Omega}_m$ is the net rate of generation, taken to be zero in our system since there is no source or sink of masses.

The sum of all the charges in the system, $-\rho = e\left(B-\sum z_m c_m \right)$ (where $B$ is the concentration of the acceptor dopant), can be related to the electric potential, $\phi$, via the Poisson's equation:
\begin{equation}
\nabla \cdot \left(\nabla \phi\right) =  -\frac{\rho}{\varepsilon}
\label{eqn:Poisson_phi}
\end{equation}
We assumed that the permittivity, $\varepsilon$, is a constant with respect to position. 
Finally, we can relate the mass flux to the carrier concentration and electric potential through the diffusion-drift equation: 
\begin{equation}
\mathbf j_m = -D_m c_m \nabla \frac{\tilde \mu_e}{k_bT} 
\label{eqn:diffusion}
\end{equation}
where $k_b$ is the Boltzmann constant and $T$ is the absolute temperature. In the dilute limit, the electrochemical potential $\mu_m$ is given by: 

\begin{equation}
\tilde \mu_m = \mu^0_m + k_b T \log\left(\frac{c_m}{c_m^0}\right) + z_m e \phi
\label{eqn:electrochem_pot}
\end{equation}

where $\mu^0_m$ is the standard chemical potential. In the steady state case and assuming that the diffusion coefficients are independent of position, combining Eq.  \ref{eqn:charge_carrier_transport} and  \ref{eqn:diffusion} yield

\begin{subequations}
\begin{eqnarray}
\nabla\cdot(c_{eon} \nabla \tilde \mu _{eon})=0 \\ 
\nabla\cdot(c_{ion} \nabla \tilde \mu _{ion})=0
\end{eqnarray}
\label{eqn:drift_diffusion_full}
\end{subequations}

and substituting $\tilde \mu_m$ with Eq.  \ref{eqn:electrochem_pot} while making the electroneutrality approximation ($B+c_{eon}- 2c_{ion}=0$) further gives

\begin{subequations}
\begin{eqnarray}
\triangle c_{eon}-\nabla c_{eon}\cdot \nabla \tilde \phi -c_{eon}\triangle\tilde \phi=0\\
\triangle c_{eon}+\nabla c_{eon}\cdot \nabla \tilde \phi+\left(B+c_{eon}\right)\triangle\tilde \phi=0
\end{eqnarray}
\label{eqns_to_solve_simplest}
\end{subequations}


where the symbol $\triangle \left(\cdot\right)$ indicates the operator $\nabla \cdot\left(\nabla (\cdot)\right)$.


\subsection{The Behavior of the Bulk}
We indicate the equilibrium quantities, such as electron and oxygen vacancy concentration, with the superscript $(0)$. In order to determine equilibrium concentrations of charge carriers, we consider the following gas phase and bulk defect reactions:
\begin{equation}
\begin{array}{lcl}
H_2({\rm gas})+O_2({\rm gas}) &\rightleftharpoons& H_2O({\rm gas}) \\
O_O^x  &\rightleftharpoons&V_O^{\bullet\bullet}+\frac{1}{2}O_2({\rm gas})+2e' 
\end{array}
\label{global_{eon}quilibrium_reactions}
\end{equation}
where the Kroger-Vink notation is used. We can also write the following equilibrium constants:

\begin{subequations}
\begin{eqnarray}
K_g = \displaystyle\frac{\tilde p_{H_2O}^2}{\tilde p_{H_2}^2\tilde p_{O_2}} \label{eqn:Kg}\\
K_r =  \displaystyle\left(\frac{c_{eon}^{(0)}}{B}\right)^2 \frac{c_{ion}^{(0)}}{B} \tilde p_{O_2}^{1/2}\\
1+ \displaystyle \frac{c_{eon}^{(0)}}{B} - 2\frac{c_{ion}^{(0)}}{B}=0
\end{eqnarray}
\label{eqns_global_reactions}
\end{subequations}
where  $\tilde p_k =\displaystyle \frac{p_k}{1 {\rm atm}}$ and $p_k$ is the partial pressure of species $k$. We solve the equilibrium concentrations of 
vacancies $c_{ion}^{(0)}$ and electrons $c_{eon}^{(0)}$ in the dilute limit at a given temperature and partial pressures.


We suppose a small bias off equilibrium is performed in our system. Experimentally, this is achieved by subjecting the electrochemical cell to a small DC voltage relative to the open circuit voltage. Alternatively, we can obtain the same information by making an AC impedance measure at open circuit and taking the resistance at the the low frequency limit where the frequency approaches zero. We indicate these small perturbations with the subscript $(1)$. These working conditions can be summarized as follows:

\begin{itemize}
\item{$c_{eon}= c_{eon}^{(0)}+c_{eon}^{(1)}$  with $\nabla c_{eon}^{(0)}=0$ and $\vert c_{eon}^{(0)}\vert\gg \vert c_{eon}^{(1)}\vert$ }
\item{$\tilde\mu_{eon}= \tilde\mu_{eon}^{(0)}+\tilde\mu_{eon}^{(1)}$ with $\nabla \tilde \mu_{eon}^{(0)}=0$ and $\vert \mu_{eon}^{(0)}\vert\gg \vert \mu_{eon}^{(1)}\vert$}
\item{$c_{ion}= c_{ion}^{(0)}+c_{ion}^{(1)}$ with $\nabla c_{ion}^{(0)}=0$ and $\vert c_{ion}^{(0)}\vert\gg \vert c_{ion}^{(1)}\vert$}
\item{ $\tilde\mu_{ion}= \tilde\mu_{ion}^{(0)}+\tilde\mu_{ion}^{(1)}$ with $\nabla \tilde \mu_{ion}^{(0)}=0$ and $\vert \mu_{ion}^{(0)}\vert\gg \vert \mu_{ion}^{(1)}\vert$}
\item{$\phi= \phi^{(0)}+\phi^{(1)}$ with $\nabla \phi^{(0)}=0$ and $\vert \phi^{(0)}\vert\gg \vert \phi^{(1)}\vert$}
\end{itemize}

Substituting the above definitions and properties above into (\ref{eqns_to_solve_simplest}) yield the following first-order relations: 

\begin{equation}
\left \{  
\begin{array} {l}
\triangle \tilde \mu_{eon}^{(1)} =0 \\ 
\triangle \tilde \mu_{ion}^{(1)} =0 
\end{array}
\right. 
\label{linearization_{eon}qns_to_solve_simplest}
\end{equation}

We remark that linearization of the electrochemical potentials around equilibrium leads to:

\begin{equation}
\left\{
\begin{array} {l}
\tilde \mu_{eon} = k_b T \log\left(\displaystyle  \frac{c_{eon}}{ c_{eon}^0}\right) -e \phi +\tilde \mu_{eon}^0 \\
\tilde \mu_{ion} = k_b T \log\left( \displaystyle \frac{c_{ion}}{ c_{ion}^0}\right) +2 e \phi +\tilde \mu_{ion}^0
\end{array}
\right. \Rightarrow
\left\{
\begin{array} {l}
\tilde \mu_{eon}^{(1)} = k_b T \displaystyle \frac{c_{eon}^{(1)}}{ c_{eon}^{(0)}}-e \phi^{(1)}  \\
\tilde \mu_{ion}^{(1)}  = k_b T\displaystyle \frac{c_{ion}^{(1)} }{ c_{ion}^{(0)}} +2 e \phi^{(1)}
\end{array}
\right.
\label{pert_electropot}
\end{equation}

Furthermore, applying the electroneutrality approximation to the equilibrium as well as the first-order perterbation in carrier concentrations gives

\begin{subequations}
\begin{eqnarray}
c_{ion}^{(0)}= \frac{B+c_{eon}^{(0)}}{2} \\
c_{ion}^{(1)} = \frac{c_{eon}^{(1)}}{2} 
\end{eqnarray}
\label{electroneutrality}
\end{subequations}

The \ref{electroneutrality} allows us to reduce Eq. \ref{pert_electropot} to a non-singular linear transformation between the variables $\left\{\phi^{(1)}, \frac{c_{eon}^{(1)}}{c_{eon}^{(0)}}\right\} $ and $\left\{\tilde\mu_{eon}^{(1)},\tilde\mu_{ion}^{(1)}\right\} $.  This implies we could use either set of unknowns to fully describe the problem. In the reminder of the paper we will use the following two quantities $n^{(1)}=  \frac{c_{eon}^{(1)}}{c_{eon}^{(0)}}$ and $\tilde \phi^{(1)} = \frac{e\phi^{(1)}}{k_b T}$.

\subsection{The Behavior of the Boundary}
Realistic boundary conditions are complex due to the formation of charge double layers \cite{Sze-81}, \cite{Rhoderick-00}. Work by Fleig et al. suggests that such electrification effects is relevant for SOFC mixed conducting cathodes \cite{Fleig-05}, \cite{ISI:000247314000018}. For simplicity, we do not consider charge double layer in our system. Furthermore, we do not consider surface diffusion as the the need to specify the surface roughness may lead to over-fitting of the data. 

As shown in Fig. 1, there are five boundaries in our electrochemical system. We start with the simplest boundary conditions. It follows from symmetry that 
$\partial_x \tilde \mu_{eon}^{(1)} =0$, $\partial_x \tilde \mu_{ion}^{(1)} =0$ on $\Gamma_2$ and $\Gamma_3$. Since the metal is ion blocking we will have $\partial_y \tilde \mu_{ion}^{(1)} = 0$  on $\Gamma_4$. By assuming that response of the metal to any perturbation is fast compared to the oxide, we can take $\phi^{(1)}$ uniform on $\Gamma_4$. Thank to linearity we can choose $ \phi^{(1)}= k_b T /e$ on $ \Gamma_4$  (so that $\tilde \phi^{(1)} = 1$) and  $ \phi^{(1)}= 0$ on $ \Gamma_1$.

The remaining boundary, $\Gamma_5$, is complex due to the gas-solid surface reaction electrochemistry. Specifically, the fuel cell anode condition under which our computation is performed requires us to consider the interaction of oxygen vacancies and electrons SDC and gas-phase hydrogen, oxygen and water vapor. A complete treatment of the surface require a detailed understanding of the electrochemical reaction pathway and kinetics parameters of various reactions that occur in series and/or parallel. However, there is little experimental data in literature regarding the surface reactions for SDC (or for any other composition of doped ceria). In the case of ceria, AC impedance spectroscopy is unable to separate multiple processes that occur on the electrode $|$ electrolyte interface due to the overwhelming "chemical" capacitance that results from redox of cerium cation between +4 and +3 oxidation states. 

In this work, we treat the surface reaction pathway as a single reaction. Specifically, we assume that the surface chemistry can be described by:
\begin{equation}
H_2({\rm gas}) \rightleftharpoons H_2O({\rm gas}) + V_O^{\bullet\bullet}+2e'
\label{global_reaction}
\end{equation}


 

Furthermore, we assume that the rate of reaction, specifically, the rate of injection of vacancies at $\Gamma_5$ satisfy
\begin{equation}
\begin{array}{lcl}
\displaystyle{\bf j}_{ion}\cdot {\bf e}_y  &=& \frac{1}{2}\displaystyle{\bf j}_{eon}\cdot {\bf e}_y  \\
\displaystyle{\bf j}_{ion}\cdot {\bf e}_y &=& k_f \tilde p_{H_2} - k_r \tilde p_{H_2O} c_{ion}c_{eon}^2\\
\label{rate_reactions}
\end{array}
\end{equation}  

where ${\bf e}_y$ is the unit vector that is perpendicular to $\Gamma_5$, $k_f$ and $k_r$ are the forward and reverse reaction rate constants, respectively. At equilibrium, the net rate of injection of both oxygen vacancies and electrons are zero, so $k_f$ and $k_r$ can be related to the equilibrium concentrations of the reactants and products: 

\begin{equation}
k_r = \frac{2 k_f \tilde p_{H_2}}{\left(c_{eon}^{(0)}+B\right) \left(c_{eon}^{(0)}\right)^2\tilde p_{H_2O}} 
\label{eq_constant}
\end{equation}

Using the same approach as Section 3.2, we compute the perturbation in the boundary condition upon applying a small bias perturbation. Combining \ref{rate_reactions} and \ref{eq_constant} and perturbing $c_{eon}= c_{eon}^{(0)}+c_{eon}^{(1)}$, we obtain: 


\begin{equation}
\begin{array}{lcl}
\displaystyle{\bf j}_{ion}^{(1)}\cdot {\bf e}_y&=& \displaystyle 2 {k_f }\left(1+\frac{c_{eon}^{(0)}}{4 c_{ion}^{(0)}}\right) \tilde p_{H_2}n^{(1)}
\end{array}
\end{equation}

We go a little further and suppose $k_f = \displaystyle 2\frac{D_{ion}}{l_c}\tilde k_f$ $\tilde k_f = \displaystyle\tilde k_f^0\tilde p_{O_2}^{\beta}\times\frac{m^3}{\#\text{particles}}$, where we choose $\beta = -1/4$.

\footnote[3]
{
The units of the $k_f$'s:
\begin{eqnarray*}
\left[k_f \right] = \frac{\# \text{particles}}{s \times m^2}\\
\left[\tilde k_f\right] = \left[ \frac{l_c}{D_{ion}} k_f\right] =  \frac{\# \text{particles}}{m^3}
\end{eqnarray*}
let us look at order of magnitude of $k_f$:
$p_{O_2}= 10^{-24}$, $l_c = 10^{-5} m$, $D_{ion} = 10^{-10} m^2/s$ and $\tilde k_f^{0} \approx 10^{32}$, so $k_f\approx10^{32}\times \frac{10^{-10}}{10^{-5}}\times10^{-6}$$=$$10^{21} \frac{\# \text{particles}}{m^2}\approx10^{-3} \frac{mol}{m^2}$$\approx 10^{-7}\frac{mol}{cm^2}$
}

It is important to note that the choice of $\beta$ is based on the oxygen partial pressure dependence of the rate limiting step(s) in the surface reaction. Since identifying the rate-limiting step in the surface reaction is beyond the scope of this work (as we described the surface reaction with a global reaction), we selected the $\beta$ value so that the $p_{O_2}$ dependence matches the experimental results \cite{ISI:000232773100001} that will be used for data analysis.

 

\subsection{Formalization of the Model}

If one defines $\mathbf x = l_c\tilde {\mathbf  x}$ ($l_c = 10 \mu m$) ,
$\displaystyle \tilde A_\phi =-\tilde k_f \frac{\tilde p_{H_2}}{c_{ion}^{(0)}}\left( 1-\frac{D_{ion}}{D_{eon}}\right )$ and $\displaystyle \tilde A_n = -\tilde k_f \frac{\tilde p_{H_2} }{c_{ion}^{(0)}}\left( 1+ 4 \frac{D_{ion} c_{ion}^{(0)}}{D_{eon} c_{eon}^{(0)}}\right)$ 
then ($\ref{linearization_{eon}qns_to_solve_simplest}$) and the boundary conditions of the previous section can be summarized as follows:
\begin{equation}
\begin{array} {l}
\triangle_{\tilde x} \tilde\phi^{(1)} =0 \\
\triangle_{\tilde x}  n^{(1)} =0
\end{array}
\label{PDE_to_solve}
\end{equation}

\begin{equation}
\left \{  \begin{array}{lclcl}
\tilde \phi^{(1)} = 0 &{\wedge}& n^{(1)} = 0& {\rm on } & \Gamma_1 \\
\partial_{\tilde x} \tilde \phi^{(1)} =0& {\wedge}&\partial_{\tilde x} n^{(1)} =0  & {\rm on }& \Gamma_2 \quad {\wedge } \quad\Gamma_3 \\
\tilde\phi^{(1)}= 1& {\wedge} &\partial_{\tilde y} n^{(1)}=-4 \frac{c_{ion}^{(0)}}{c_{eon}^{(0)}}\partial_{\tilde y} \tilde \phi^{(1)} & {\rm on } & \Gamma_4 \\
\partial_{\tilde y} \tilde \phi^{(1)} = \tilde A_\phi n^{(1)}& {\wedge} &\partial_{\tilde y} n^{(1)} =\tilde A_n n^{(1)}&{\rm on } &  \Gamma_5
\end{array}
\right.
\label{BC_to_solve}
\end{equation}

\subsubsection{Numerical Method}
In order to solve numerically the Equations (\ref{PDE_to_solve}) with the boundary conditions  (\ref{BC_to_solve}) we employ an h-adapted finite element method (FEM). FEM is well known for elliptic problem such as the one we are studying, see for example \cite{Fichera-65} and \cite{Quarteroni-94}. 

In order to employ FEM appropriately, we first recast the problem in the following weak form, where $m$ and $\psi$ are test functions on the domain $\Omega$:
\begin{equation}
\begin{array}{lcl}
\displaystyle\int_{\Omega} \nabla m\cdot \nabla n^{(1)}\ud \mathcal A+4\frac{c_{ion}^{(0)}}{c_{eon}^{(0)}}\int_{\Gamma_4}m \partial_{\tilde y} \tilde\phi^{(1)} \ud \gamma-\tilde A_n\int_{\Gamma_5} m n^{(1)} \ud \gamma &=&0 \\
\displaystyle\int_{\Omega} \nabla \psi\cdot \nabla \phi^{(1)}\ud \mathcal A-\tilde A_\phi\int_{\Gamma_5}\psi n^{(1)} \ud \gamma &=&0
\end{array}
\label{weak_eon_eqns}
\end{equation}

with the addition of the conditions that:
\begin{equation}
\left\{
\begin{array}{lclcl}
\tilde \phi^{(1)} &=& 0 \quad\wedge\quad n^{(1)} =1&\text{on}&\Gamma_1 \\
\tilde \phi^{(1)} &=&1 &\text{on}&\Gamma_4
\end{array}
\right.
\end{equation}

The discrete version of equations (\ref{weak_eon_eqns}) are then solved using FreeFem++ \cite{hec-pir-2007}. The equations are initially discretized on a triangular unstructured mesh, using quadratic continuous basis functions with a third order bubble. The mesh is adaptively refined up to seven times at each solution step and the a posteriori adaptation is performed against $\tilde \mu_{eon}^{(1)}$. The h-adaptation ensures high regularity of the $H^1$ a posteriori estimator \cite{Brenner-00}, locally below $0.01\%$,  and it guarantees that the mesh is finer where the sharpest gradients occur. We note that mesh adaptivity results is coarseness everywhere except in the vicinity of the interfaces, in particular the refinement increases as we approach the triple phase boundary; this fact indicates strong nonlinearities around that area. Eleven integral tests were also implemented in order to ensure that at solution step the numerical method is consistent with the boundary conditions and it globally satisfies conservation of charge.
Finally we note that FreeFem++ execution time is comparable to custom-written C++ code and its speed his enhanced by the utilization of fast direct linear solvers such as the multi-frontal package UMFPACK \cite{992206}. Due to the sparsity of the problem we make extensive use of this last feature. 


\subsection{Value of the Polarization Resistance}
To compute the electrode polarization resistance, let us first consider the relevant electrochemical currents that exist within our system. Due to the mixed conducting nature of ceria, there will be an inherent cross-plane electronic current, termed $I_e^{CP}$, that flows between the metal current collectors located on the opposite side of ceria (Figure \ref{fig:current_schematics}). The surface reaction taking place on $\Gamma_5$ will simultaneously inject one oxygen vacancy and two electrons into ceria. Two distinct current result: the cross-plane ionic current $I_g^{CP}$ that flows between the two sides of the cell, and the in-plane electron drift-diffusion current $I_g^{IP}$ that flows between gas $|$ ceria interface and the metal $|$ ceria interface. By electroneutrality, $I_g^{CP}$ = $I_g^{IP}$. We formally define these currents as follows:

\begin{equation}
\begin{array}{l}
\displaystyle I_e^{CP} = \int_{\Gamma_1} \mathbf j_{eon}\cdot \mathbf n \ud x = \int_{\Gamma_4} \mathbf j_{eon}\cdot \mathbf n \ud x \\
\displaystyle  I_g^{IP} = \int_{\Gamma_5} \mathbf j_{eon}\cdot \mathbf n \ud x = \int_{\varphi^S_e(\Gamma_5)} \mathbf j_{eon}\cdot \mathbf n \ud x 
\end{array}
\end{equation}

Note that to obtain $I_g^{IP}$, we could integrate the current density either over $\Gamma_5$ or over $\varphi^S_e(\Gamma_5)$, which, as depicted Fig. \ref{fig:Topology}, represents some fraction of $\Gamma_4$ accessed by the current injected from the gas $|$ ceria interface. Direct comparison with the work of Jamnik and Maier, \cite{ISI:000168035300017}, and Lai and Haile, \cite{ISI:000232773100001}, leads to the following definitions of the bulk electronic resistance $R_{eon}$, bulk ionic resistance $R_{ion}$, and the electrode polarization resistance normalized by the sample area $R_{ion}^\perp$: 

\begin{equation}
\begin{array}{l}
\displaystyle  R_{eon} = 2\frac{<\tilde \mu_{eon}^\star>_{\Gamma_4}-<\tilde \mu_{eon}^\star>_{\Gamma_1}}{j_e}=2\frac{<\tilde \mu_{eon}^\star>_{\Gamma_4}}{j_e}\\
\displaystyle  R_{ion} = 2\frac{<\tilde \mu_{eon}^\star>_{\Gamma_5}-<\tilde \mu_{eon}^\star>_{\Gamma_1}}{j_g} =2\frac{<\tilde \mu_{eon}^\star>_{\Gamma_5}}{j_g} \\
\displaystyle  R_{ion}^\perp = \frac{<\tilde \mu_{eon}^\star>_{\Gamma_4}-<\tilde \mu_{ion}^\star>_{\Gamma_5}}{j_g}
\end{array}
\label{eqn:defn_Resistances}
\end{equation}

where $j_{e} = \frac{I_e^{CP}}{W_1+W_2}$ and $j_{g} = \frac{I_g^{IP}}{W_1+W_2}$ are the current densities averaged over the total sample area. 
Our two-dimensional model allows us to separate various contributions of $R^\perp_{ion}$. We shall discuss two physically significant separations of the contributions. First, it is possible to separate $R^\perp_{ion}$ into a surface reaction resistance term $R_{surf} $ and in-plane electron drift-diffusion term $R_{eon-DD} $. $R_{surf} $ corresponds to the electron-vacancy electrochemical potential difference at the gas $|$ ceria interface, and $R_{eon-DD} $ corresponds to the electron potential difference at the gas $|$ ceria interface and at the metal $|$ ceria interface. 

\begin{equation}
\begin{array}{l}
\displaystyle R_{surf} =   \frac{< \tilde\mu_{eon}^\star>_{\Gamma_5}-< \tilde\mu_{ion}^\star>_{\Gamma_5}}{I_g} = \left( 1+\frac{\bar n}{4 \bar p}\right)\frac{<n^{(1)}>_{\Gamma_5}}{j_g}\\
R_{eon-DD} = R^\perp_{ion} - R_{eon-ion}
\end{array}
\label{eqn:defn_Rsurf}
\end{equation}

Physically speaking, $R_{surf} $ represents the resistance associated with the chemical transformation of electrons to vacancy at the gas $|$ ceria chemical reaction site. Specifically, $R_{surf} $ is due to the migration of oxygen vacancy from the bulk to the surface, and the subsequent chemical reactions that give rise to the electronic current. Finally, the drift-diffusion of the injected electrons from the reaction site to the metal current collector results in $R_{eon-DD}$. 

Alternatively, we could also separate $R^\perp_{ion}$ into a "true" polarization term, $R_{pol}$, and a "deviation" term, $R_{avg} $, that results from averaging:

\begin{equation}
\begin{array}{l}
\displaystyle R_{pol} =   \frac{<\tilde \mu_{eon}^\star>_{\varphi_e^S(\Gamma_5)}-< \tilde\mu_{ion}^\star>_{\Gamma_5}}{j_g}\\
R_{avg} = R^\perp_{ion} - R_{pol}
\end{array}
\label{eqn:defn_Rpol}
\end{equation}

In Eq. (\ref{eqn:defn_Resistances}), $R^\perp_{ion}$ is proportional to the difference of the electronic electrochemical potential averaged over the metal $|$ ceria and averaged over gas $|$ ceria interface. In Eq. (\ref{eqn:defn_Rpol}), we define the $R_{pol}$ by averaging only some portion of  $\Gamma_4$ (rather than over the entire interface) by considering the interface mapped by current lines injected from gas $|$ ceria interface. $R_{avg} $, defined as the difference between the electrode polarization resistance and the true electrode polarization, is simply a spurious contribution due to averaging.

\label{model}

\section{Results}
\subsection{Potential Distributions and Surface Regions}

Electrochemical equipotential lines for oxygen vacancies (Fig. \ref{fig:pretty}, right) calculated using various values for the surface reaction rate constant, $\tilde k_f^0$, reveal that the potential and current distribution exhibit a relatively weak dependence on $\tilde k_f^0$. In general, oxygen vacancy equipotential lines bend as they approach the oxygen vacancy blocking metal$|$ceria interface ($\Gamma_4$) from the bulk. On the other hand, equipotential lines for electrons (Fig. \ref{fig:pretty}, left), display substantial deviations from those for oxygen vacancies, due to the presence of two current sources: cross-plane electronic current that flows between the current collectors on opposite side of the cell, and the in-plane electronic current injected by the surface reaction that flows between the metal$|$ceria ($\Gamma_4$) and the gas$|$ceria ($\Gamma_5$) interface. The electron potential distributions also depend strongly on the magnitude of $\tilde k_f^0$, indicating that electronic current injected from the surface reaction taking place at ($\Gamma_5$) strongly influence the electron penetration depth of the so-called "surface region."  

\begin{figure}[ht]
 	\begin{center}
 		\scalebox{0.3}{\includegraphics{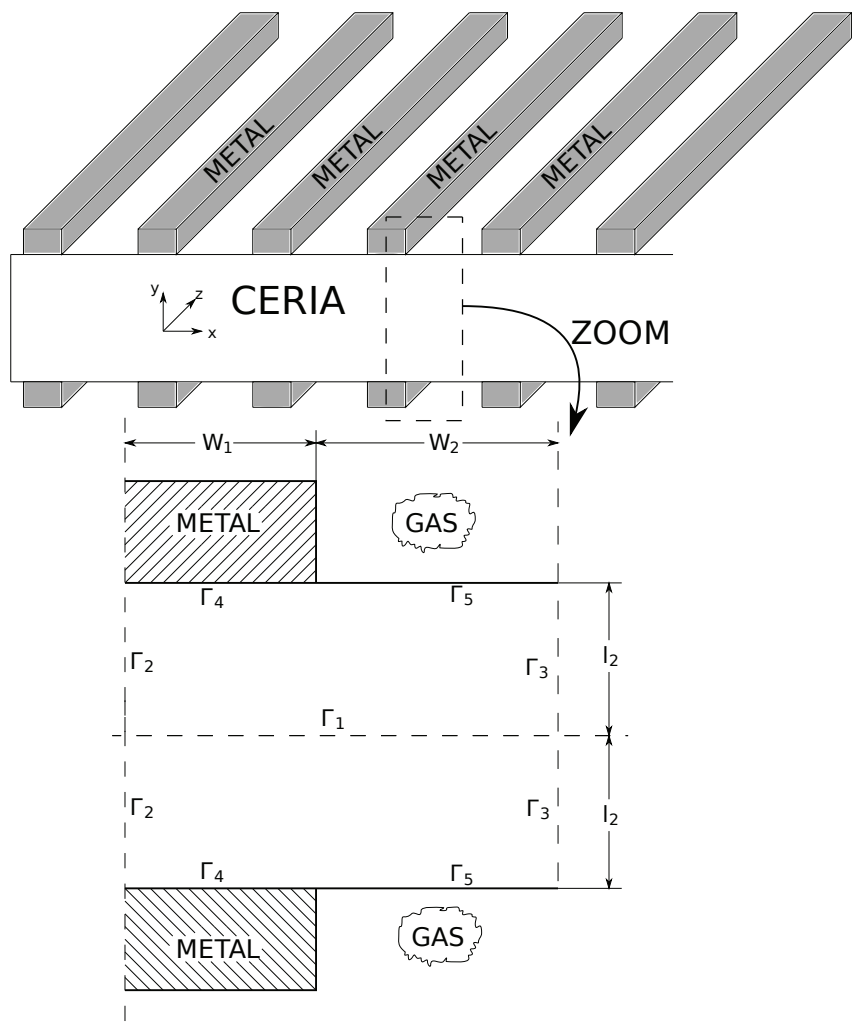}}
		\caption{top: A symmetric cell with patterned Pt stripes on both sides of dense ceria placed in a uniform gas environment.
		bottom: Schematic depiction of the boundaries. $\Gamma_1$, $\Gamma_2$, and $\Gamma_3$ are symmetry lines, while $\Gamma_4$ is the metal $|$ ceria interface, and $\Gamma_5$ is the gas $|$ ceria interface. 2$W_1$ is the width of the metal, 2$W_2$ is the width of the ceria directly exposed to the gas phase, and 2$l_2$ is the thickness of ceria.}
		\label{fig:Mathematical_domain}
  	\end{center}
\end{figure}

\begin{figure}[ht]
 	\begin{center}
 		\scalebox{0.7}{\includegraphics{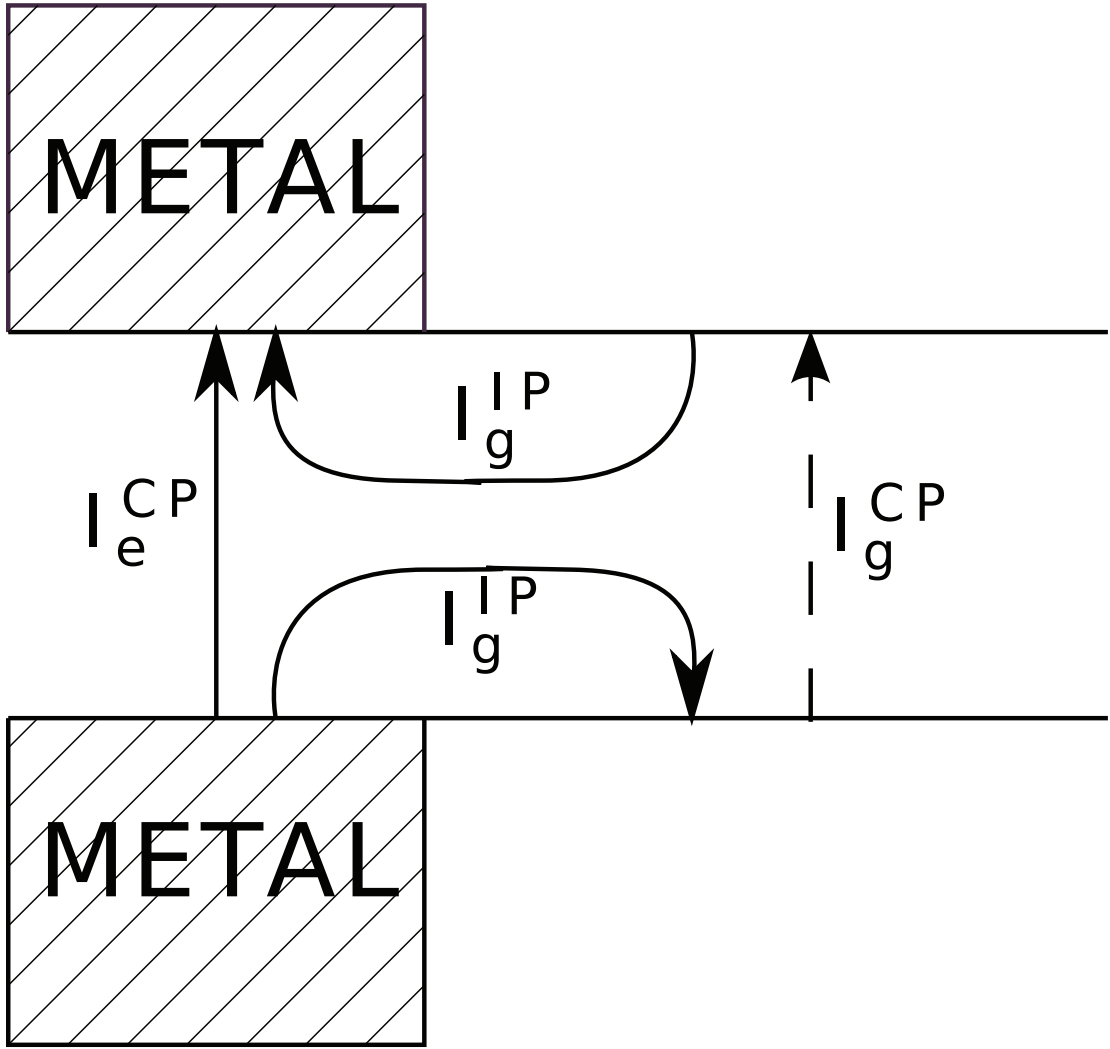}}
		\caption{Various electronic and ionic current within the cell. Solid line indicates the electronic current and dashed line indicates the oxygen vacancy (ionic) current. The superscript ÒIPÓ refers to the in-plane current that flows between the gas $|$ ceria chemical reaction site and the metal current collector, while ÒCPÓ refers to the cross-plane current that flows between the metal current collectors located on the opposite side of ceria. }
		\label{fig:current_schematics}
  	\end{center}
\end{figure}

\begin{figure}[ht]
 	\begin{center}
 		\scalebox{0.7}{\includegraphics{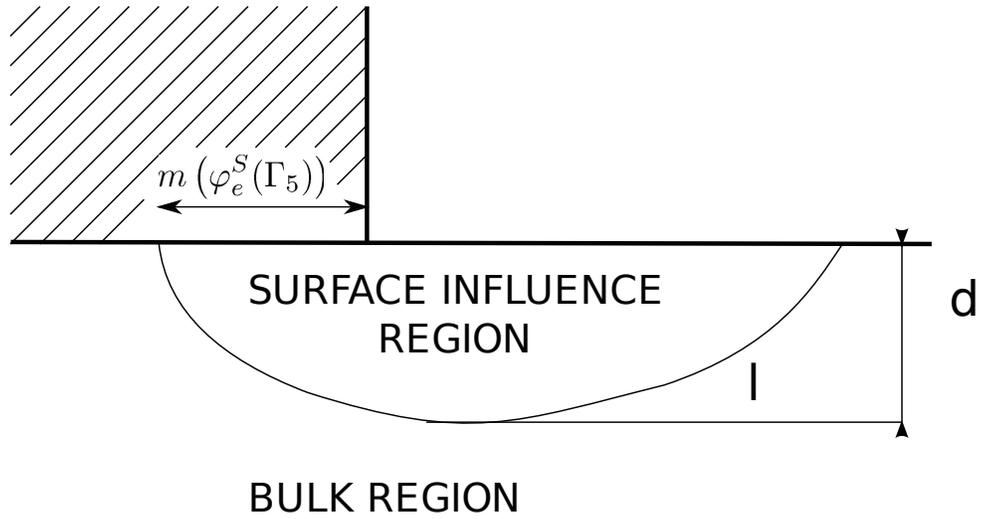}}
		\caption{Illustration of the Òsurface regionÓ, where the in-plane electronic drift-diffusion current prevails. The dimension of the region is indicated by the length $l$ anddepth $d$. $\varphi^S_e(\Gamma_5)$ depicts the fraction of the metal $|$ ceria interface mapped by the electronic current injected from gas $|$ ceria interface.}
		\label{fig:Topology}
  	\end{center}
\end{figure}

Fig.~\ref{fig:boundary_influence_oxidizing} shows the boundary of the surface region, given by the "trajectory" of electrons injected from the surface reaction site furthest from the metal current collector (the intersection of $\Gamma_3$ and $\Gamma_5$  in Fig.~\ref{fig:Mathematical_domain}). Physically, the surface zone can be viewed as a region where electronic current is entirely the in-plane electronic current ($I_g^{IP}$), rather than the cross-plane current $I_e^{CP}$. The surface region dimensions (Fig.s \ref{fig:Topology}, \ref{fig:boundary_influence_oxidizing}, \ref{fig:topology_650}) are specified by the largest length ($l$), largest depth ($d$), and the area ($A$). All dimensions increases as a function of $\tilde k_f^0$. As the penetration area increases, the in-plane electrons will flow through a larger cross-section of ceria, thereby reducing the diffusion resistance. It is interesting to note that the surface region approaches an asymptote for large $\tilde k_f^0$, suggesting that when surface reactions are sufficiently fast, i.e. when they are in electrochemical equilibrium, the total electron injection current will be dominated by the in-plane electron diffusion resistance. 

The surface region dimensions also grow with increasing $\tilde p_{O_2}$, though it is more pronounced for higher $\tilde k_f^0$. At lower $\tilde k_f^0$ values, the penetration area is virtually independent of $\tilde p_{O_2}$. The penetration depth is a function of the relative magnitude of $I_e^{IP}$ to $I_e^{CP}$. As the ratio $I_e^{IP}/I_e^{CP}$ grows, for instance, as a function of $\tilde p_{O_2}$, the penetration depth is expected to increase. In Fig.~\ref{fig:topology_650}, we see that an increase in the penetration dimensions is indeed accompanied by an increase in $I_e^{IP}/I_e^{CP}$.

\subsection {Electrode Polarization Resistance}
 
Eq. (\ref{eqn:defn_Rpol}) states that the electrode polarization resistance, $R^\perp_{ion}$, can be expressed as a sum of the true polarization term $R_{pol}$ and a deviation term $R_{avg}$ that results from averaging the electrochemical potential of electrons across the entire metal $|$ ceria interface $\Gamma_4$ rather than just the region accessed by the in-plane electronic current $\varphi^S_e(\Gamma_5)$ (Fig.~\ref{fig:Topology}). We examine the extent of deviation of the $R^\perp_{ion}$ from $R_{pol}$ by computing:
\begin{eqnarray}
f_{pol} = \frac{R_{pol}} {R_{ion}^\perp}\\
\end{eqnarray}
Under a variety of conditions, $f_{pol}$ is very close to unity (Fig.~\ref{fig:splits_Rpol_Rioneon} top), indicating that the deviation term is quite small compared to the true polarization resistance. For the remainder of our analysis, we approximate $R_{pol}$ = $R^\perp_{ion}$.

\subsection {Electron Diffusion Resistance}

In Eq. (\ref{eqn:defn_Rsurf}), the total electrode polarization resistance, $R_{ion}^\perp$, is expressed as a sum of a surface reaction resistance term, $R_{surf}$, and an in-plane electron diffusion resistance term, $R_{eon-DD}$. Under the moderately reducing $\tilde p_{O_2}$ regime (where the electron carrier concentration is negligible compared to the extrinsic oxygen vacancies formed by acceptor-doping), $R_{eon-DD}$ is proportional to approximately $\tilde p_{O_2}^{1/4}$ (Fig. \ref{fig:Rioneon_Reoneon} bottom), following the same $\tilde p_{O_2}$ dependence as bulk electronic resistivity. As for $R_{surf}$, it is also proportional to $\tilde p_{O_2}^{1/4}$ as a result of our choice of $\beta$ (Fig. \ref{fig:Rioneon_Reoneon} top). Turning to the dependence on $\tilde k_f^0$, we observe that both $R_{surf}$ and $R_{eon-DD}$ decreases with increasing $\tilde k_f^0$. However, a significant difference between $R_{eon-DD}$  and $R_{surf}$ is that, in a log-log plot the former approaches an asymptotic value as a function of $\tilde k_f^0$, whereas the latter does not. This interesting behavior of  $R_{eon-DD}$ is directly related with the asymptotic behavior of the the penetration depth of electrons injected from $\Gamma_5$ to $\Gamma_4$ (and vice-versa) as a function of $\tilde k_f^0$ (Fig. \ref{fig:boundary_influence_oxidizing}). To help us examine $I_g^{IP}$, we further define fractional surface reaction resistance and fractional electron drift-diffusion as:
\begin{equation}
\begin{array}{l}
f_{surf} = \frac{R_{surf}} {R_{ion}^\perp}\\
f_{eon-DD} = 1 - f_{surf}
\end{array}
\end{equation}

Plotting $f_{surf}$ as a function of $\tilde k_f^0$ (Fig. \ref{fig:splits_Rpol_Rioneon} bottom) reveals that when the surface reaction is very fast (i.e. large $\tilde k_f^0$) , $f_{surf}$ approaches zero and $R_{ion}^\perp$ is dominated by $R_{eon-DD}$. On the other hand, when the surface reaction is slow, $f_{surf}$ approaches unity and $R_{ion}^\perp$ is dominated by $R_{surf}$, as would be expected. When considering only material property dependencies (i.e. neglecting $\tilde p_{O_2}$, $T$, and sample geometry) , $R_{surf}$ is only a function of $\tilde k_f^0$ whereas $R_{eon-DD}$ is a function of both $\tilde k_f^0$ and $\sigma_{eon}$. As $\tilde k_f^0$ tends toward infinity, $R_{surf}$ approaches zero and $R_{eon-DD}$ approaches an asymptotic limit that is a function of only $\sigma_{eon}$. In other words, as the surface reaction resistance term becomes negligible, electron carrier concentration and mobility alone determines the penetration dimensions and therefore $R_{ion}^\perp$. The condition under which $f_{surf}$ approaches zero corresponds to the physical case where the electrode reaction is limited by the rate in which the electrons migrate from the gas $|$ ceria reaction site to the metal rather than the rate of surface reaction. Generally speaking, for a wide-bandgap mixed conductor exhibiting a low or moderate electronic conductivity and high $\tilde k_f^0$, such as ceria, in-plane electron drift-diffusion cannot be neglected. Accordingly, the electron diffusion length, (separation between the metal in Fig.~\ref{fig:Mathematical_domain} top), needs to be tuned in order to minimize the electrode polarization resistance. 

\subsection {Topological Considerations} 

There are two degrees of freedom in the metal current collector topology: the metal stripe width ($2W_1$) and the intermetal distance ($2W_2$). Fig.~\ref{fig:iso_var_low_kf} show parametric plots of the the fractional surface resistance (top row) and the total electrode polarization resistance (normalized for the total sample area) (middle row) and as a function of $W_1$,$\displaystyle\frac{W_2}{W_1}$, and $\tilde k_f^0$ at select temperatures and $\tilde p_{O_2}$. We observe the general trend that increasing $\displaystyle\frac{W_2}{W_1}$ (gas $|$ ceria interface to metal $|$ ceria interface ratio) leads to a reduction in the polarization resistance. Specifically, under the conditions that the fractional surface reaction resistance is greater than 0.99, we observe a linear decrease in the polarization resistance with increasing $\displaystyle\frac{W_2}{W_1}$. When the electron drift-diffusion resistance is negligible, the current density of electrons injected from $\Gamma_5$ is essentially uniform as a function of the position (and distance to the metal), and thus the fraction of area available for electrochemical surface reaction, given by $\displaystyle f = \frac{W_2}{W_1+W_2}$, determines the polarization resistance. Fig. ~\ref{fig:iso_var_low_kf} (bottom row) shows the polarization resistance normalized by the gas $|$ ceria interface area and confirms that the normalized resistance remains relatively constant as long as the electron diffusion fractional resistance is negligible. However, as the fractional surface resistance decreases (due to an increase in $\tilde k_f^0$, for instance), electrochemical surface reactions taking place closer to the metal will inject a larger current into the oxide. For example, at $\tilde k_f^0 = 7.5\times 10^{33}$, $\tilde p_{O_2} = 4.1\times 10^{-26}$ and $T = 650^oC$, a significant nonlinearity as well as a distinct minima in the polarization resistance as a function of $W_1$ and $W_2$ (Fig.~\ref{fig:iso_var_low_kf}  is observed (top right)). Furthermore, Fig. ~\ref{fig:iso_var_low_kf} 
(lower right) shows the polarization resistance normalized for the gas $|$ ceria interface area begins to deviate from the constant values, confirming parts of the interface is becoming less active due to increased electron diffusion resistance at spatial positions further away from the metal current collector. 

In general, when the surface reaction rate constant is small or when the bulk electronic conductivity is large, one should increase $f$ in order to increase maximize the area available for surface reactions, as long as the electron diffusion fractional resistance is kept low. On the other hand, when the surface rate rate constant is large or when the bulk electronic conductivity is small, one needs to find an intermediate $f$ in order to balance the area available for surface reaction and the in-plane electron diffusion distance. 
	
\subsection{Comparison to Experimental Results}

We fit the polarization resistance data obtained by Lai and Haile \cite{ISI:000232773100001} using AC impedance spectroscopy on a cell geometry consistent with our model description. The experimental result was based on a $\text{porous Pt} | SDC | \text{porous Pt}$  cell in $\text{H}_2-\text{H}_2\text{O}-\text{Ar}$. We approximated the porous Pt electrode as line patterns by estimating $W_1$ and $W_2$ based on the actual pore size and interpore distance. We fit the polarization resistance using $\tilde k_f^0$ as the only parameter and fixed dopant and equilibrium carrier concentration according to the values obtained in the experiment. It should be noted that all parameters were obtained from the same electrochemical cell by Lai and Haile and are highly self-consistent. The fitting (Fig.~\ref{fig:opti_plot}) shows computed $\tilde k_f^0$ corresponding to the polarization resistance obtained experimentally. Because we phenomenologically set the $\tilde p_{O_2}$ dependence of $R_{surf}$ to 1/4 so that $R_{ion}^\perp$ would exhibit the same $\tilde p_{O_2}$ as the experimental data (and the other component of the polarization resistance, $R_{eon-DD}$, is also proportional to $\tilde p_{O_2}^{1/4)}$ in the same way as the bulk electronic conductivity), obtaining the same dependence in $\tilde p_{O_2}$ for the experimental and fitted value is automatic. 

Taking the fitted $\tilde k_f^0$ values, we can further separate the polarization resistance into the surface reaction and the electron drift-diffusion contributions. At the temperatures and $\tilde p_{O_2}$ examined, the computed $f_{surf}$ (Fig.~\ref{fig:vary_splits_Rioneon_Rperp}) is close to unity (for $W_1\approx 1.5 \mu m $ and $W_2/W_1\approx 1.67$), implying that the surface reaction step is the rate-limiting step. To examine the dependence of $f_{surf}$ on the geometric parameter (which directly influences the electron diffusion length and area of the gas $|$ ceria interface), we fit $\tilde k_f^0$  to the polarization resistances while varying $W_1$ and $W_2$ . The parametric plot (Fig.~\ref{fig:vary_splits_Rioneon_Rperp}) again shows that  $f_{surf}$ is close to unity for a wide range of $W_1$, $W_2$, $T$ and $\tilde p_{O_2}$. However, we do observe the general trend that $f_{surf}$ decreases slightly with increasing $W_1$ and decreasing  $W_2/W_1$. Decreasing $W_2/W_1$ (at a fixed $W_1$) reduces the electron diffusion length and reduces the area of the gas $|$ ceria interface, and $\tilde k_f^0$ needs to be increased in order to fit to the observed polarization resistance (Fig. \ref{fig:opti_vary_W1_over_W2}). For the same reason discussed in Section 4.3, this leads to a decrease in $f_{surf}$. On the other hand, increasing $W_1$ (at a fixed $W_2/W_1$) increases the electron diffusion length without affecting the available reaction area. As a result, increased $R_{eon-DD}$ leads to an decrease in $f_{surf}$. 

Approximating a grid-like porous metal on ceria as line patterns could lead to some errors, such as over-estimating the fraction of gas $|$ ceria interface and the electron diffusion length. However, given that the computed $f_{surf}$ is far from 0.5 (the case where surface reaction and electron drift-diffusion are equally co-limiting) for a wide range of  $W_1$  and $W_2$, these errors will not change the $f_{surf}$  significantly and will only re-scale the magnitude of the resistances slightly. Therefore, based on the numerical analysis in this work, the electrode reaction in $\text{porous Pt} | SDC | \text{porous Pt}$  cell in $\text{H}_2-\text{H}_2\text{O}-\text{Ar}$ is likely to be surface reaction limited. 

Finally, it should be noted that our assumption that the electron mobility and equilibrium carrier concentration is the same in the near-surface region and in the bulk directly determines the contribution of the in-plane electron drift-diffusion resistance to the electrode polarization resistance. Since electron penetration depth is predicted to be on the order of $1 \mu m$, dopant segregation and presence of blocking grain boundaries near the surface could in principle affect the local electron mobility and concentration.

\label{kinetics}

\section{Conclusions}
A two-dimensional electrochemical model has been developed for mixed conductors with patterned metal current collectors. Numerical simulation for a Pt $|$ SDC $|$ Pt in reducing atmosphere revealed a strong nonlinearity in the electronic potential and current distributions near the surface. In particular, we show that the in-plane electron drift-diffusion current plays a crucial role in determining the surface electrochemical behavior. Under certain conditions, the in-plane electron drift-diffusion resistance could dominate the electrode resistance. 
\label{kinetics}

\section{Acknowledgments}
The authors gratefully acknowledge financial support for this 
work by the Office of Naval Research under grant N00014-05-1-0712. \\
The authors thank Prof. Fr\'ed\'eric Hecht for his valuable insight and support on 
Freefem++. 

\newpage
\appendix
\section{Definitions}
Some nomenclature:
\begin{enumerate}
\item{$c_{eon} =$ concentration of electrons in $\#\text{particles}/m^3$}
\item{$c_{ion} =$ concentration of vacancies in $\#\text{particles}/m^3$}
\item{$B =$ concentration of negatively charged background particles in $\#\text{particles}/m^3$}
\item{$\mu_{eon} =$ chemical potential of electrons $= \log\left(\frac{c_{eon}}{c_{eon}^0}\right)+\mu_{eon}^0$}
\item{$\mu_{ion} =$ chemical potential of vacancies $= \log\left(\frac{c_{ion}}{c_{ion}^0}\right)+\mu_{ion}^0$}
\item{$\tilde\mu_{eon} =$ electrochemical potential of electrons $= \mu_{eon} - e \phi$}
\item{$\tilde\mu_{ion} =$ electrochemical potential of vacancies $= \mu_{ion} +2 e \phi$}
\item{$\tilde\mu_{eon}^\star =$ star electrochemical potential of electrons $=\phi-\frac{1}{e}\mu_{eon}$}
\item{$\tilde\mu_{ion}^\star =$ star electrochemical potential of vacancies $=\phi+\frac{1}{2e} \mu_{ion} $}
\item{$\phi =$ electric potential}
\item{$\mathbf j_{eon}=$electron current $=e D_{eon} c_{eon} \nabla  \frac{\tilde \mu_{eon}}{k_B T}$}
\item{$\mathbf j_{ion} =$ vacancy current $=-2e D_{eon} c_{eon} \nabla \displaystyle  \frac{\tilde \mu_{eon}}{k_B T}$}
\item{$\mathbf j_{eon}^P=$ electron flux$ =- D_{eon} c_{eon} \nabla  \frac{\tilde \mu_{eon}}{k_B T}$}
\item{$\mathbf j_{ion}^P=$ vacancy flux$ =- D_{ion} c_{ion} \nabla  \frac{\tilde \mu_{ion}}{k_B T}$}
\item{$U_T = \displaystyle \frac{k_b T}{e}$}
\end{enumerate}


\section{Assumptions}
For our Pt $|$ SDC $|$ Pt cell, we make the following assumptions:

\begin{enumerate}
\item{the conditions are steady-state, i.e. $\partial_t(\cdot)=0$ for all unknowns;}
\item{electroneutrality is satisfied (i.e. $c_{eon}-2c_{ion}+B({\bf x}) \approx 0$) throughout the sample;}
\item{the drift-diffusion equations describe correctly the fluxes;}
\item{diffusivities of electrons and vacancies are constant and uniform;}
\item{surface diffusion is frozen;}
\item{the chemistry is correctly described in one global step.}
\end{enumerate}


\clearpage




\bibliographystyle{pccp.bst}	
\bibliography{refs.blb}		


\clearpage

\begin{table}
\begin{center}
\caption{\label{tab1}Insert table caption here}
\begin{tabular}{ccccc}
\hline
$T$&$ 500^oC$ & $550^oC$ & $600^oC$ &$650^oC$ \\ \hline
$u_{eon}$ $\left[ \frac{m^2}{ V^2 s}\right]$  &$4.762\times10^{-8}$ &$6.257\times10^{-8}$ &$6.873\times10^{-8}$ & $8.123\times10^{-8}$\\
$u_{ion}$ $\left[ \frac{m^2}{ V^2 s}\right]$& $1.166\times10^{-9}$ & $2.070\times10^{-9}$ & $3.359\times10^{-9}$ & $4.936\times10^{-9}$\\
$K_g$ & $5.059\times10^{27}$& $4.814\times10^{25}$ & $7.757\times10^{23}$ & $1.944\times10^{22}$ \\
$K_r$ & $5.008\times10^{-22}$& $2.263\times10^{-20}$ & $6.610\times10^{-19}$ & $1.340\times10^{-17}$  \\
\hline
&$W_1~[\mu m]$&$W_2 ~[\mu m]$ & $L ~[\mu m]$ & $B\cite{Andersson03072006} \cite{zhan:A427} ~\left[ \frac{\text{\#particles}}{m^3}\right]$ \\ \hline
&$1.5 $ & $2.5$ & $500$ & $3.47 \times 10^{27}$\\
\hline
\label{table:diffusivity}
\end{tabular}
\footnotetext[1]{from }
\end{center}
\end{table}


\clearpage

\begin{list}{}{\leftmargin 2cm \labelwidth 1.5cm \labelsep 0.5cm}

\item[\bf Fig. 1] Caption of Figure 1.
\item[\bf Fig. 2] Caption of Figure 2.

\end{list}

\clearpage

\clearpage

\clearpage

\clearpage

\begin{figure}[ht]
 	\begin{center}
 		\scalebox{0.6}{\includegraphics{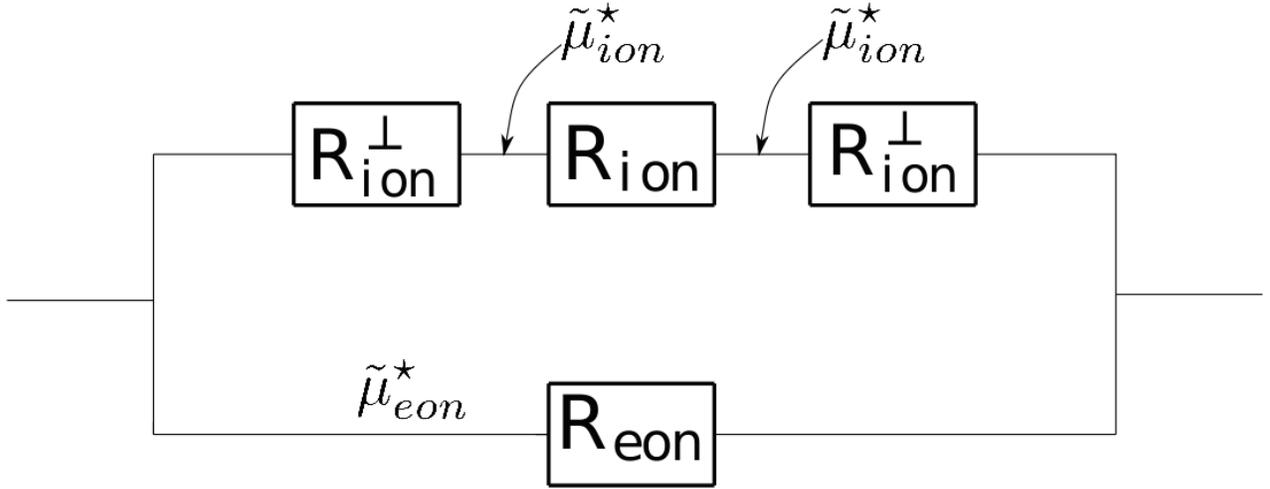}}
		\caption{Simplified one-dimensional equivalents circuit employed by Lai and Haile, and Jamnik and Maier, where $R_{eon}$ is the bulk electronic resistance, $R_{ion}$ is the bulk ionic (oxygen vacancy) resistance, and $R^\perp_{ion}$ is the electrode polarization resistance normalized by the cell area. $\tilde\mu_{ion}^\star$ and $\tilde\mu_{eon}^\star$ are the electrochemical potential of oxygen vacancies and electrons, respectively.}
		\label{fig:circuit}
  	\end{center}
\end{figure}

\clearpage

\begin{figure}[ht]
 	\begin{center}
 		\scalebox{0.7}{\includegraphics{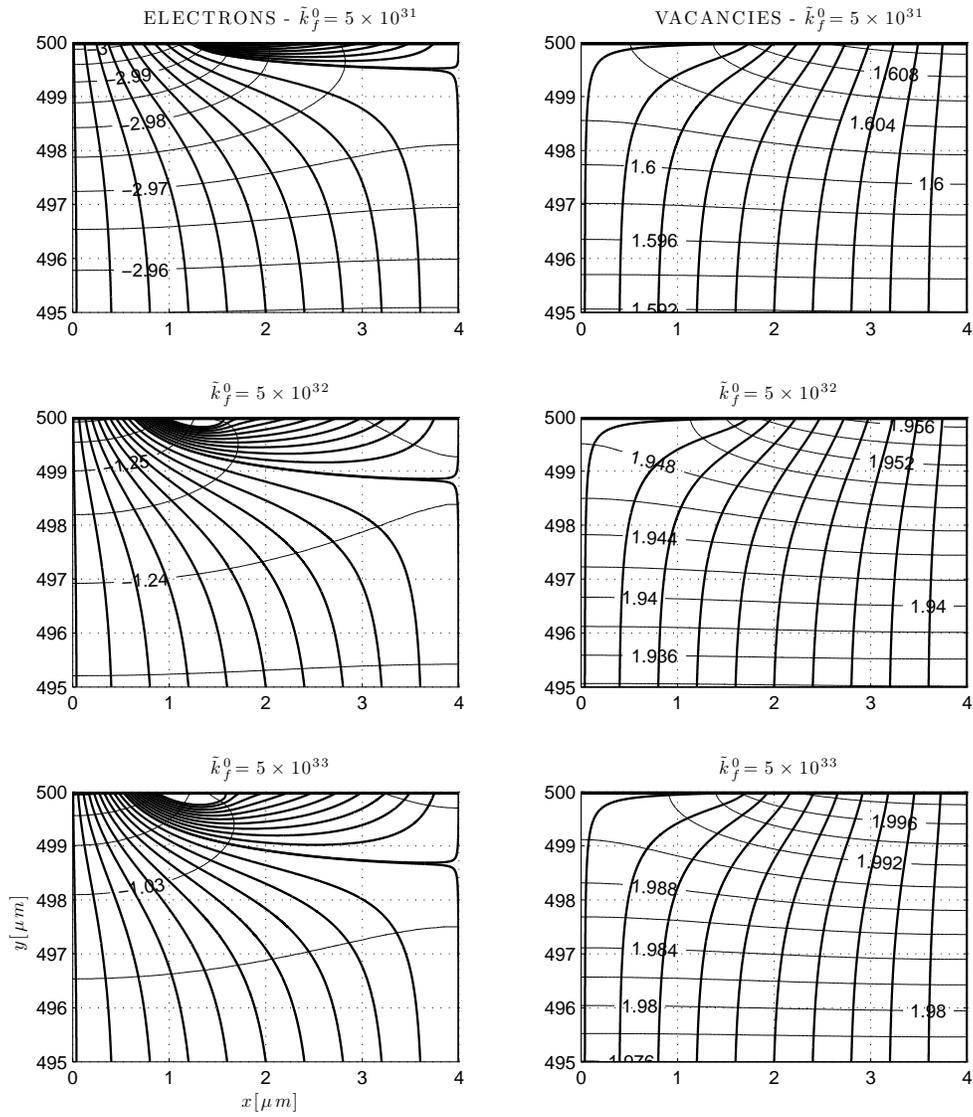}}
		\caption{Electrochemical equipotential lines (left) and the corresponding current flow lines (right) computed for various surface reaction rate constants kf0 at 650C, pO2 = 4.1E-26 atm.}
		\label{fig:pretty}
  	\end{center}
\end{figure}



\clearpage

\begin{figure}[ht]

 	\begin{center}

 		\scalebox{0.9}{\includegraphics{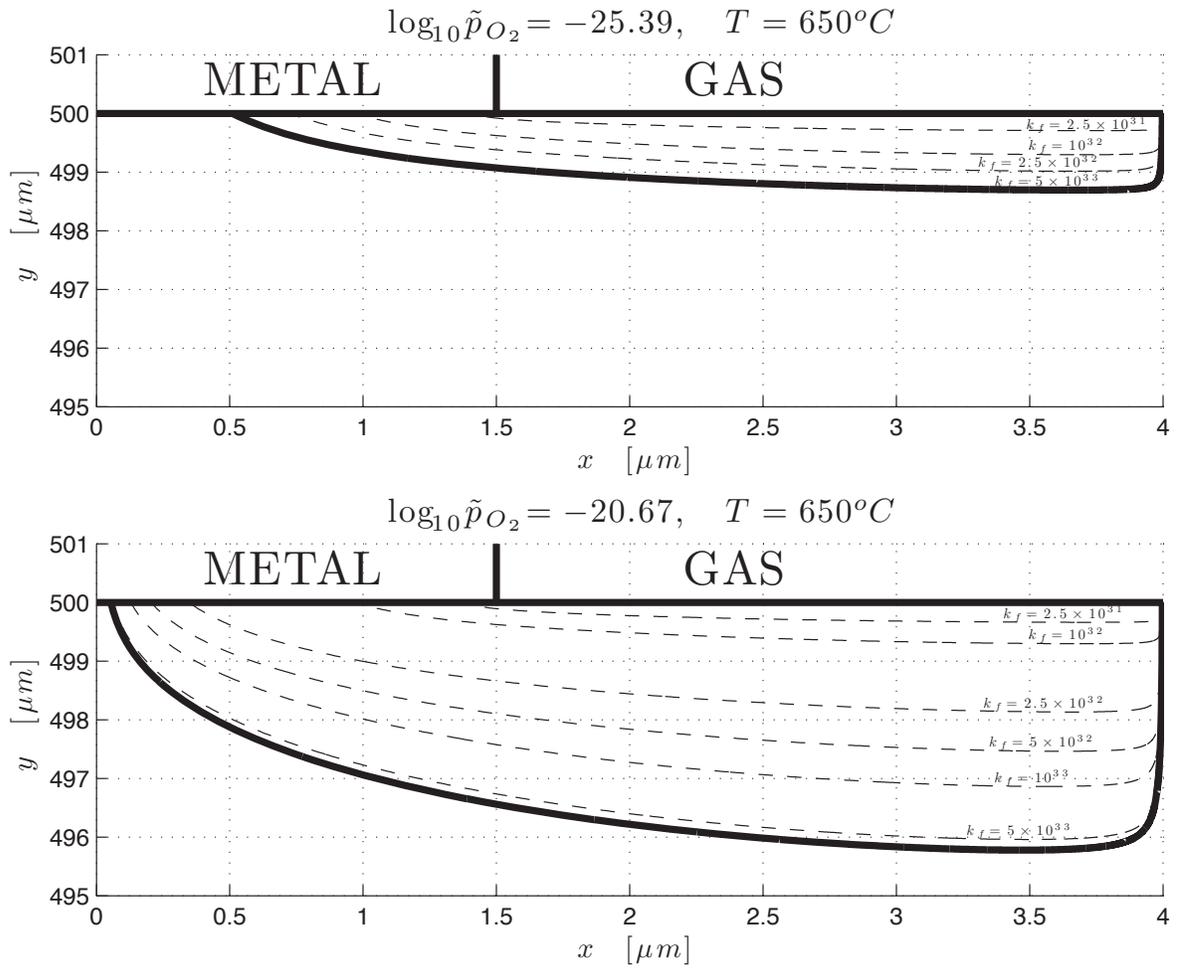}}

		\caption{The boundary of the surface region, where electrons undergo in-plane drift-diffusion between the gas $|$ ceria inteface and the metal current collectors, computed for various surface reaction rate constants kf0 at 650C, pO2 = 4.1E-26 atm (top) and 2.1E-21 atm (bottom).}
	\label{fig:boundary_influence_oxidizing}

  	\end{center}

\end{figure}


\clearpage

\begin{figure}[ht]
 	\begin{center}
 		\includegraphics{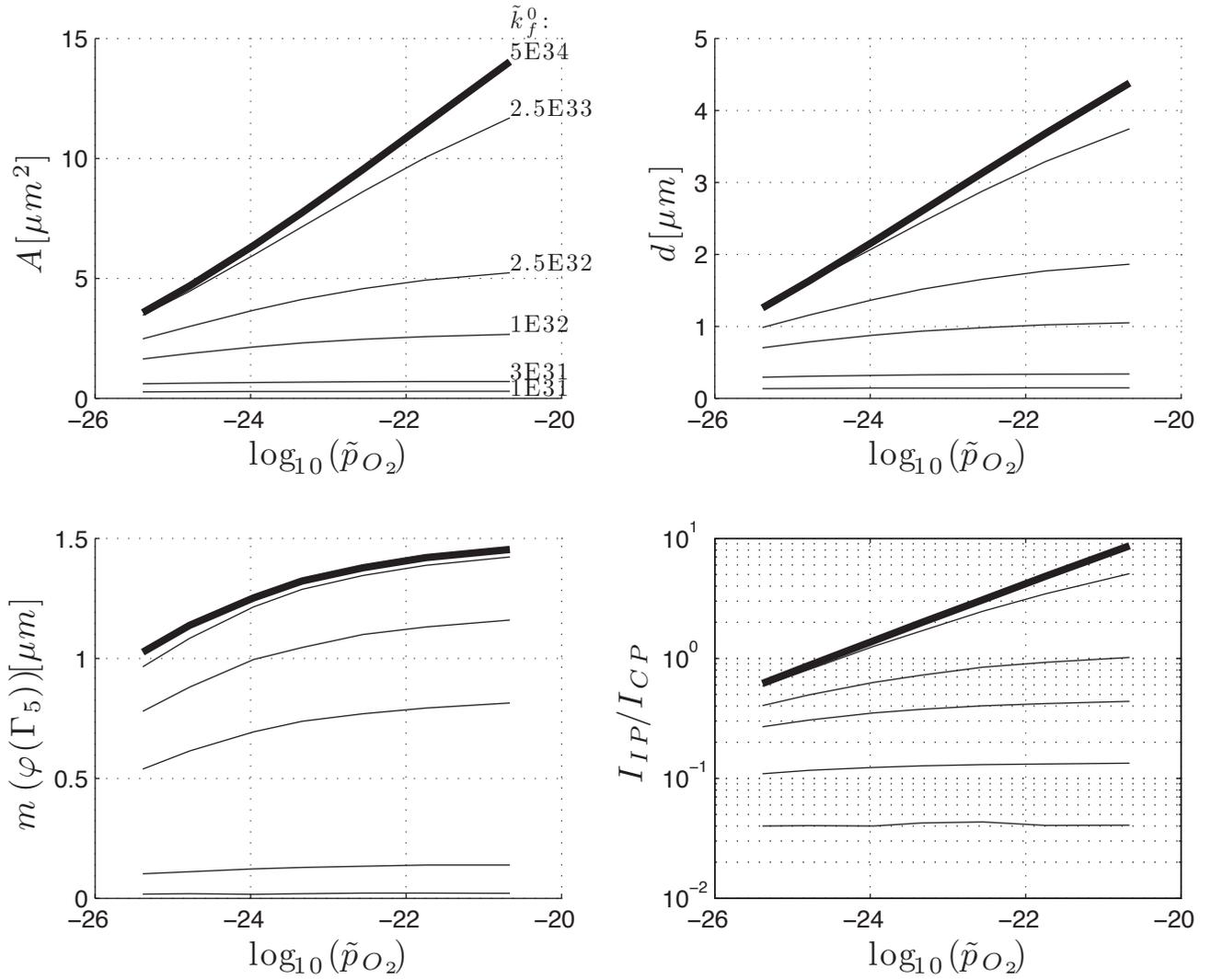}
		\caption{Dimensions of the surface region as a function of $\tilde p_{O_2}$ and $\tilde k_f^{(0)}$ at $650^oC$.}
		\label{fig:topology_650}
  	\end{center}
\end{figure}

\clearpage
\begin{figure}[ht]
 	\begin{center}
 		\scalebox{0.7}{\includegraphics{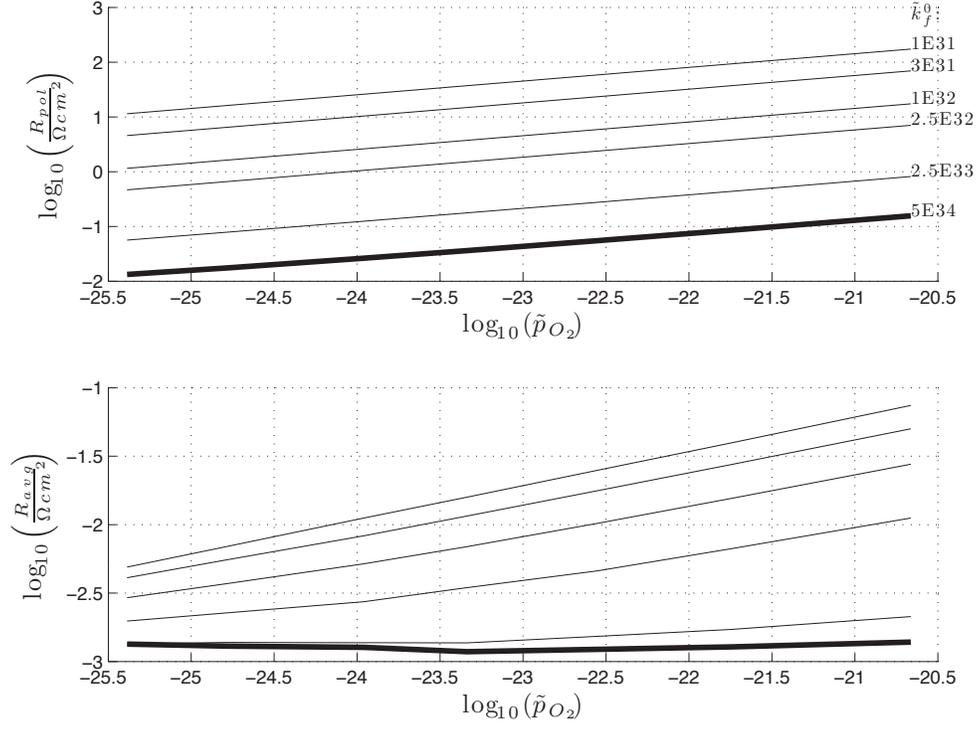}}
		\caption{Absolute value of the true polarization resistance $R_{pol}$ (top) and the deviation term $R_{avg}$ (bottom) as a function of pO2 and $\tilde k_f^{(0)}$ at $650^oC$.}
		\label{fig:Rpol_Ravg}
  	\end{center}
\end{figure}

\clearpage

\begin{figure}[ht]
 	\begin{center}
 		\scalebox{0.7}{\includegraphics{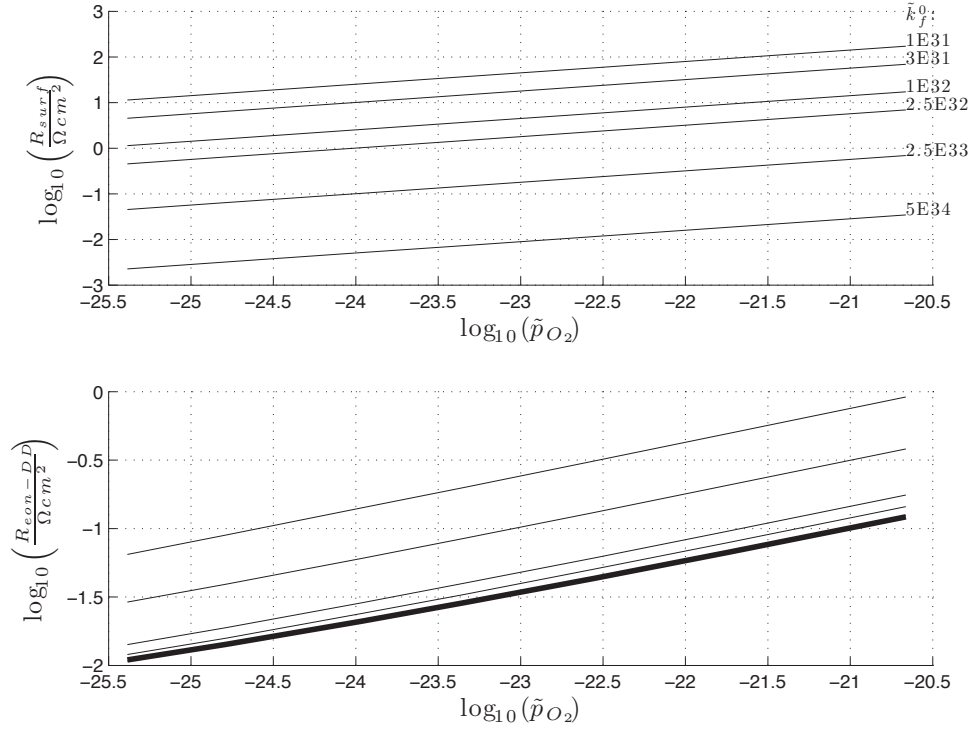}}
		\caption{Absolute value of the surface reaction resistance $R_{surf}$ (top) and the electron drift-diffusion resistance $R_{eon-DD}$ as a function of pO2 and $\tilde k_f^{(0)}$ at $650^oC$.}
		\label{fig:Rioneon_Reoneon}
  	\end{center}
\end{figure}

\clearpage
\begin{figure}[ht]
 	\begin{center}
 		\includegraphics{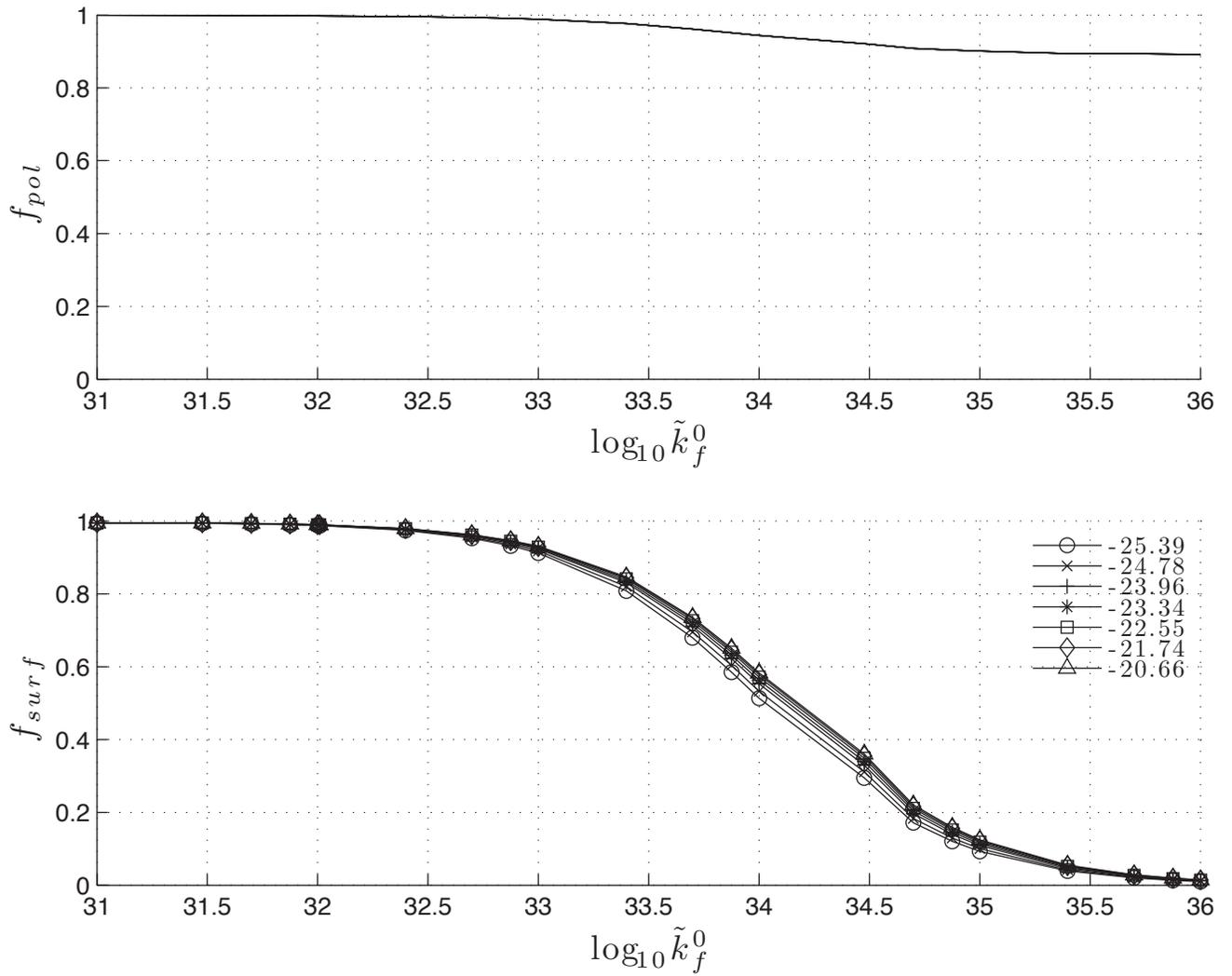}
		\caption{Fractional true polarization resistance (top) and fractional surface reaction resistance (bottom) as a function of $\tilde k_f^{(0)}$ at $650^oC$, parametrized with respect to $\log_{10}\tilde p_{O_2}$.}
		\label{fig:splits_Rpol_Rioneon}
 	\end{center}
\end{figure}


\clearpage
\begin{figure}[ht]
 	\begin{center}
 		\includegraphics{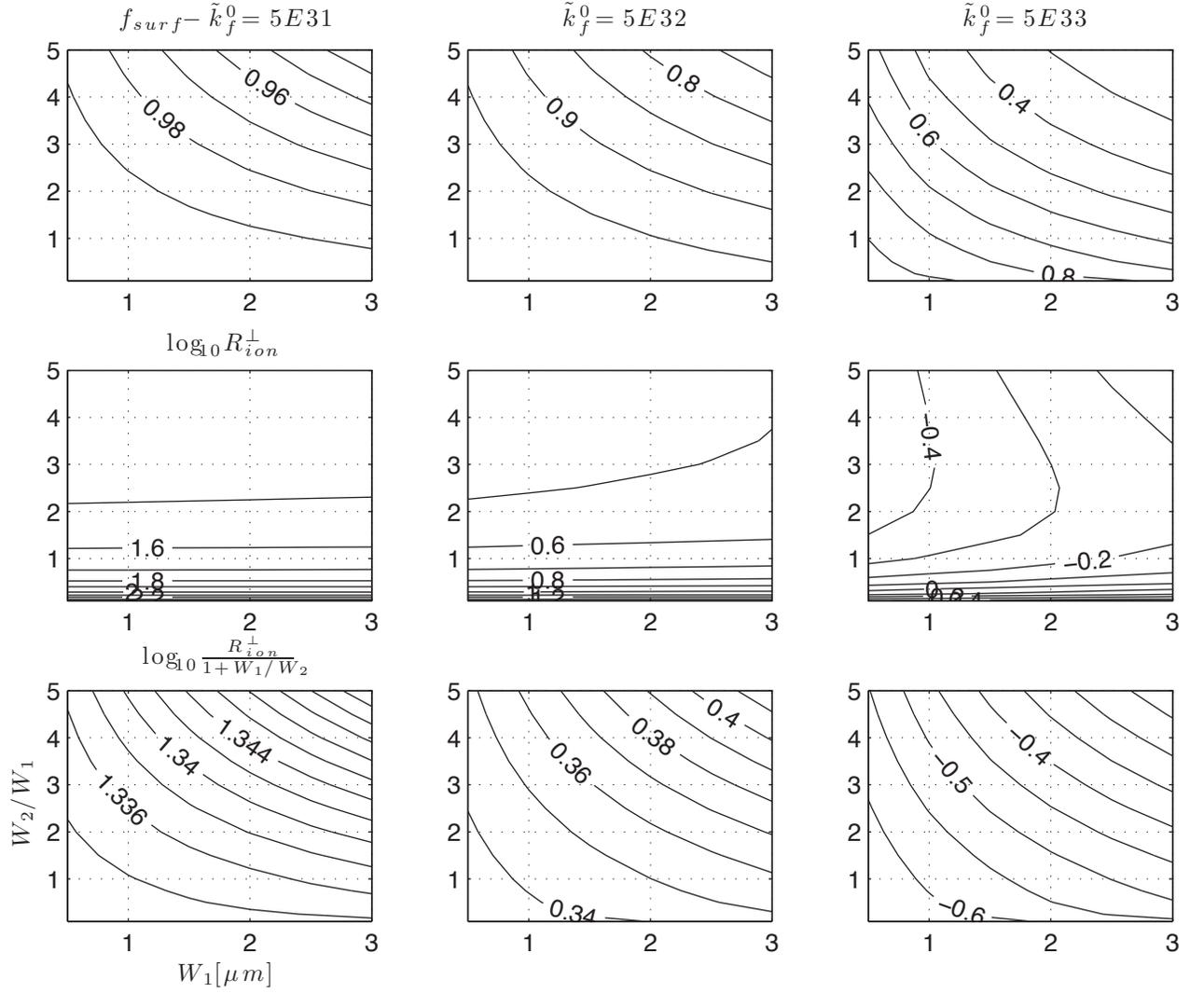}
		\caption{Parametric plots of the fractional surface reaction resistance (top row), the electrode polarization resistance $R_{ion}^\perp$ normalized for the sum of the metal $|$ ceria and gas $|$ ceria interfacial area (middle row), and normalized for the gas $|$ ceria interfacial area (bottom row), as a function of $W_2/W_1$ and $W_1 $}
		\label{fig:iso_var_low_kf}
  	\end{center}
\end{figure}

\clearpage
\begin{figure}[ht]
 	\begin{center}
 		\includegraphics{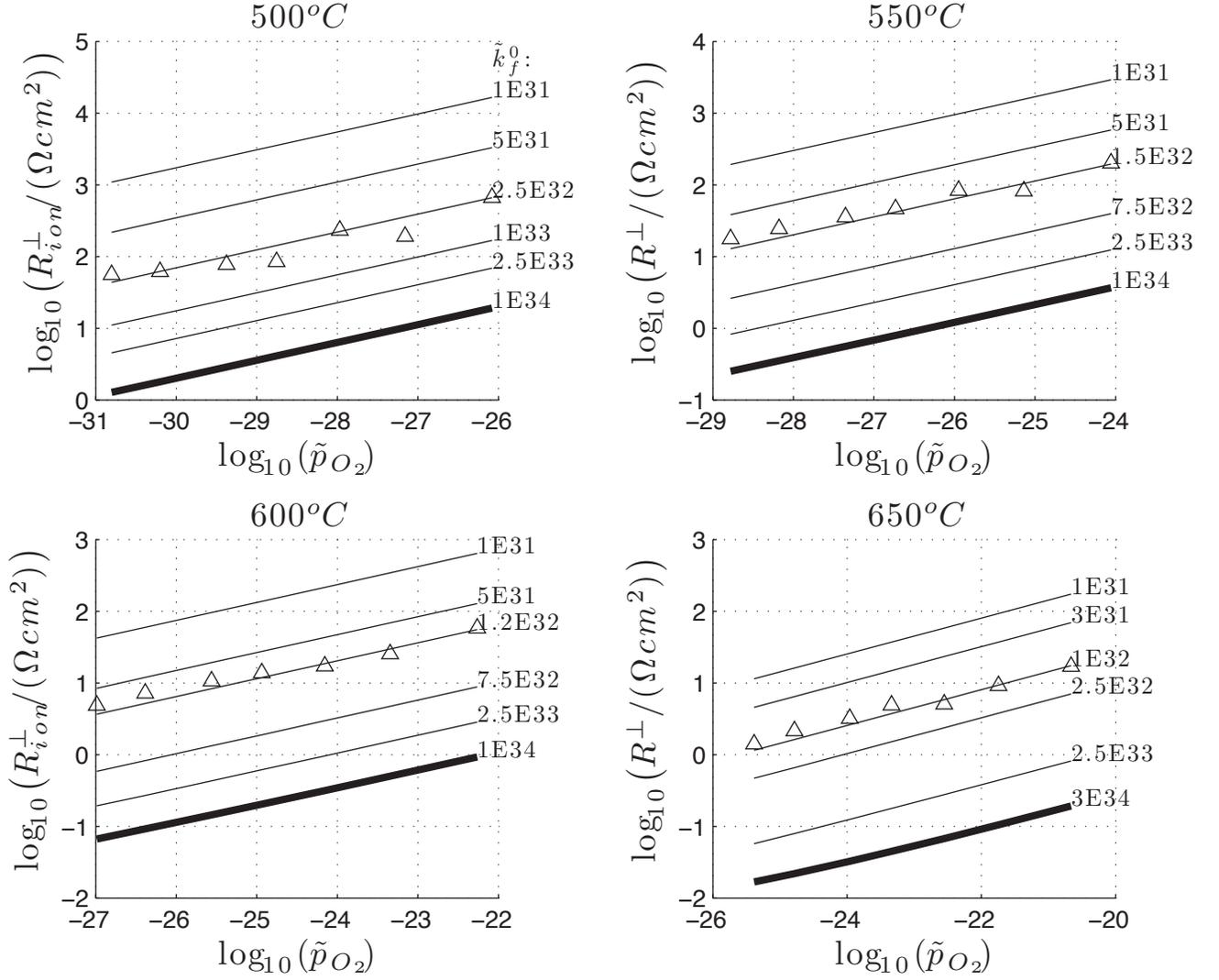}
		\caption{Electrode polarization resistance , $R_{ion}^\perp$, plotted as a function of $\tilde p_{O_2}$ and $\tilde k_f^{(0)}$ at $500^oC$, $550^oC$, $600^oC$ and $650^oC$. The open triangles show the experimental data obtained in ref \cite{ISI:000252830600012}.}
		\label{fig:opti_plot}		
  	\end{center}
\end{figure}


\clearpage
\begin{figure}[ht]
 	\begin{center}
 		\includegraphics{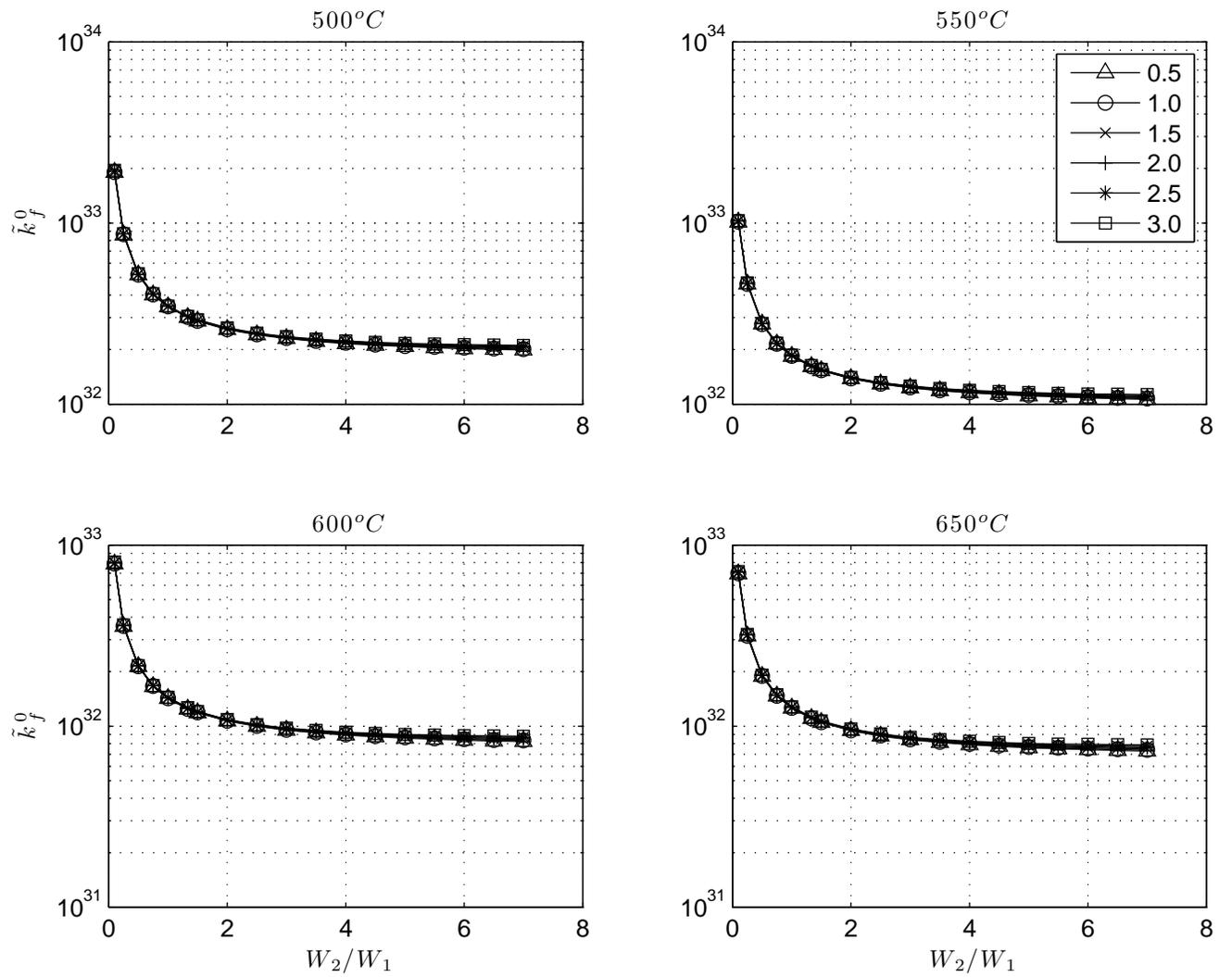}
		\caption{If we assume that that our initial choice of $W_1$ and $W_2$ is not the correct one, it is legitimate to ask the following question: "which $\tilde k_f^0$ fit the ASRP data best?". We find that the fitting depends only on the ratio $\frac{W_2}{W_1}$ and not on the chosen value of $W_1$, the label indicates that $W_1 =0.5, 1.0, 1.5, \ldots 3.0 \mu m$; subsequent pictures show that the Area affected by the surface $\Gamma_5$ goes with $W_1^2$. This is a check of how good the numerical study is.}
		\label{fig:opti_vary_W1_over_W2}
  	\end{center}
\end{figure}

\clearpage





\clearpage


\clearpage

\begin{figure}[ht]
 	\begin{center}
 		\scalebox{0.7}{\includegraphics{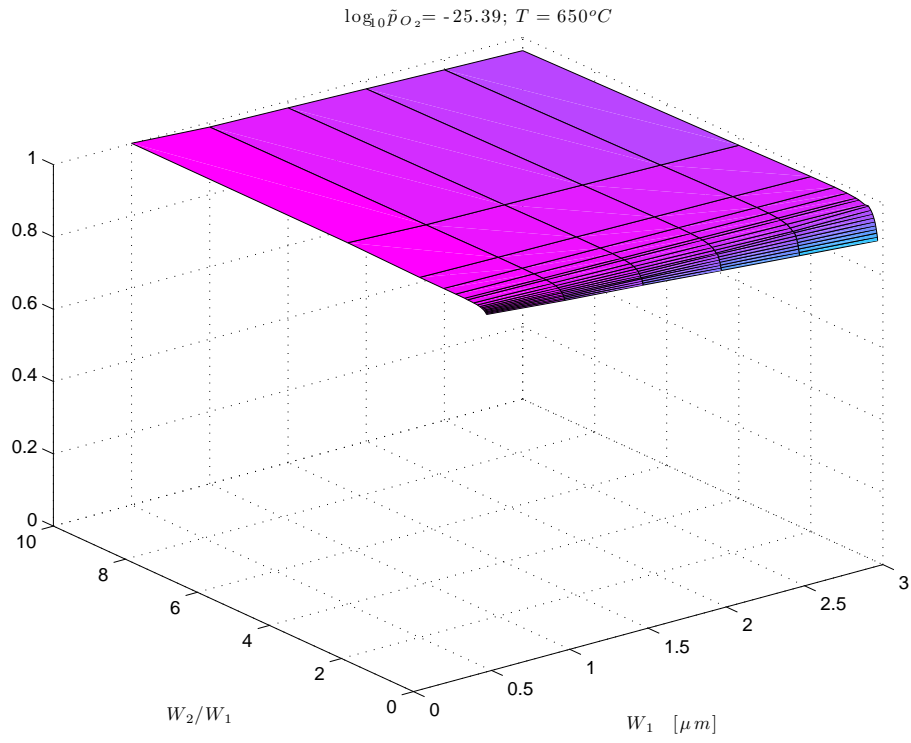}}
		\caption{Fractional surface reaction resistance, obtained after fitting $\tilde k_f^{(0)}$ to the experimental data in ref ???, and plotted as a function of pO2 $\tilde k_f^{(0)}$ at $500^oC$, $550^oC$, $600^oC$ and $650^oC$}
		\label{fig:vary_splits_Rioneon_Rperp}
  	\end{center}
\end{figure}

\clearpage

\end{document}